\documentclass{article}

\usepackage{arxiv}

\usepackage[utf8]{inputenc} 
\usepackage[T1]{fontenc}    
\usepackage[hidelinks]{hyperref}      
\usepackage{url}            
\usepackage{booktabs}       
\usepackage{amsfonts}       
\usepackage{nicefrac}       
\usepackage{microtype}      
\usepackage{graphicx}
\usepackage{doi}
\usepackage{float}

\title{Total heat flux convergence in the calculation of 2d and 3d heat losses through building elements}

\date{} 					

\author{ \href{https://orcid.org/0000-0003-4534-8441}{\includegraphics[scale=0.06]{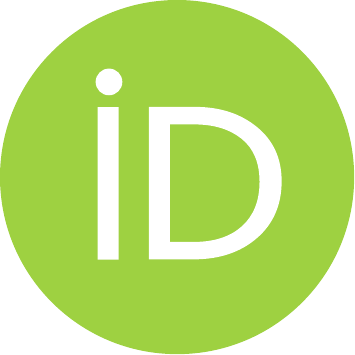}\hspace{1mm}Sanjin Gumbarević}\thanks{Submitted to Fourteenth International Conference on Computational Structures Technology \& Eleventh International Conference on Engineering Computational Technology 2021} \\
	University of Zagreb\\
	Faculty of Civil Engineering\\
	Department of Materials\\
	\texttt{sgumbarevic@grad.hr} \\
	\And
	\href{https://orcid.org/0000-0003-0264-2958}{\includegraphics[scale=0.06]{orcid.pdf}\hspace{1mm}Bojan Milovanović} \\
	University of Zagreb\\
	Faculty of Civil Engineering\\
	Department of Materials\\
	\texttt{bmilovanovic@grad.hr} \\
	\AND
	\href{https://orcid.org/0000-0002-9100-4060}{\includegraphics[scale=0.06]{orcid.pdf}\hspace{1mm}Mergim Gaši} \\
	University of Zagreb\\
	Faculty of Civil Engineering\\
	Department of Materials\\
	\texttt{mgasi@grad.hr} \\
	\And
	Marina Bagarić\\
	University of Zagreb\\
	Faculty of Civil Engineering\\
	Department of Materials\\
	\texttt{mbagaric@grad.hr} \\
}

\renewcommand{\shorttitle}{ECT2021}

\hypersetup{
pdftitle={Total heat flux convergence in the calculation of heat losses through building elements},
pdfauthor={Sanjin Gumbarević,Bojan Milovanović, Mergim Gaši, Marina Bagarić},
pdfkeywords={Finite Element Method, Steady-state heat conduction, Numerical analysis, Thermal bridges, Code\_Aster},
}

\begin{document}
	
\maketitle

\begin{abstract}
	Heat losses through the building envelope is one of the key factors in the calculation of the building energy balance. If heat conduction through building elements that form the building envelope is analysed, then the heat flux can be examined as one, two or three-dimensional. If steady-state heat conduction is observed, which is commonly used to assess the heat losses in building, there is an analytical solution for one-dimensional problem. For two and three-dimensional problems, especially for the complex geometry cases, one must use numerical methods to solve the heat conduction equation. To standardise the calculation of the two and three-dimensional calculation of heat losses through building elements (described with steady-state heat conduction), ISO 10211:2017 standard can be used. The standard has four benchmark examples with criteria that must be satisfied to declare a method as a high-precision calculation method for the calculation of thermal bridges. A problem occurs for Case 1 of the standard’s benchmark test because the analysed problem has a singular point due to discretely assigned Dirichlet boundary conditions. The reliability of the results around the singular point could be improved by the refinement of the mesh in the area around the singular point, but as a point of interest is the total heat flux that is entering the building element, and it must converge between two subdivisions, this method is not good since the reliable result cannot be reached. The problem for the convergence is in the marginal node because the temperature gradient in it increases as the temperature difference remains constant and the distance between the corresponding nodes decreases, and that difference does not change as the number of subdivisions changes from 2n × 2n to 	n × n subdivisions, as defined in the standard. One could argue on the necessity of this benchmark test because there is no such building detail (with singular point due to discretely assigned boundary conditions) in reality. Even if there is a discontinuity in temperature field on the boundary, there is an interval in which this change is to happen, and the heat flux has a theoretical limit which is not infinity but rather some finite number. From the results of this research, it is shown that one should neglect a certain number of singular points in order to achieve the tolerance given by the standard since the temperature further from the marginal node is stable for any subdivision.
\end{abstract}

\keywords{Finite Element Method \and Steady-state heat conduction \and Numerical analysis \and Thermal bridges \and Code\_Aster}

\section{Introduction}

With increasing global awareness on climate change, building energy analysis came into focus in both expert and scientific work. Heat losses through the building envelope are one of the key factors in building energy balance. Depending on the position of a construction detail in the building envelope, heat transfer can be one, two or three-dimensional. \ref{e1} is the equation which describes the unsteady state heat conduction, where $c$ is heat capacity, $S$ is a source or sink term, and $q$ is the Fourier’s law of heat conduction (\ref{e2}).

\begin{equation} \label{e1}
	c \, \frac{\partial T}{\partial t} = - \, \textit{div} \, \textbf{q} + S
\end{equation}

\begin{equation} \label{e2}
	\textbf{q} = grad \, (-k \, T)
\end{equation}

When observing the steady-state problem, no heat is accumulated inside the element, so the part that contains the volumetric heat capacity can be neglected. \ref{e3} is the equation that describes this problem (where $k$ is the thermal conductivity), and it has an analytical solution for one-dimensional heat transfer. 

\begin{equation} \label{e3}
	- \, div \, [grad \, (-k \, T)] = 0
\end{equation}

\begin{figure}[H]
	\centering
	\includegraphics[scale=0.06]{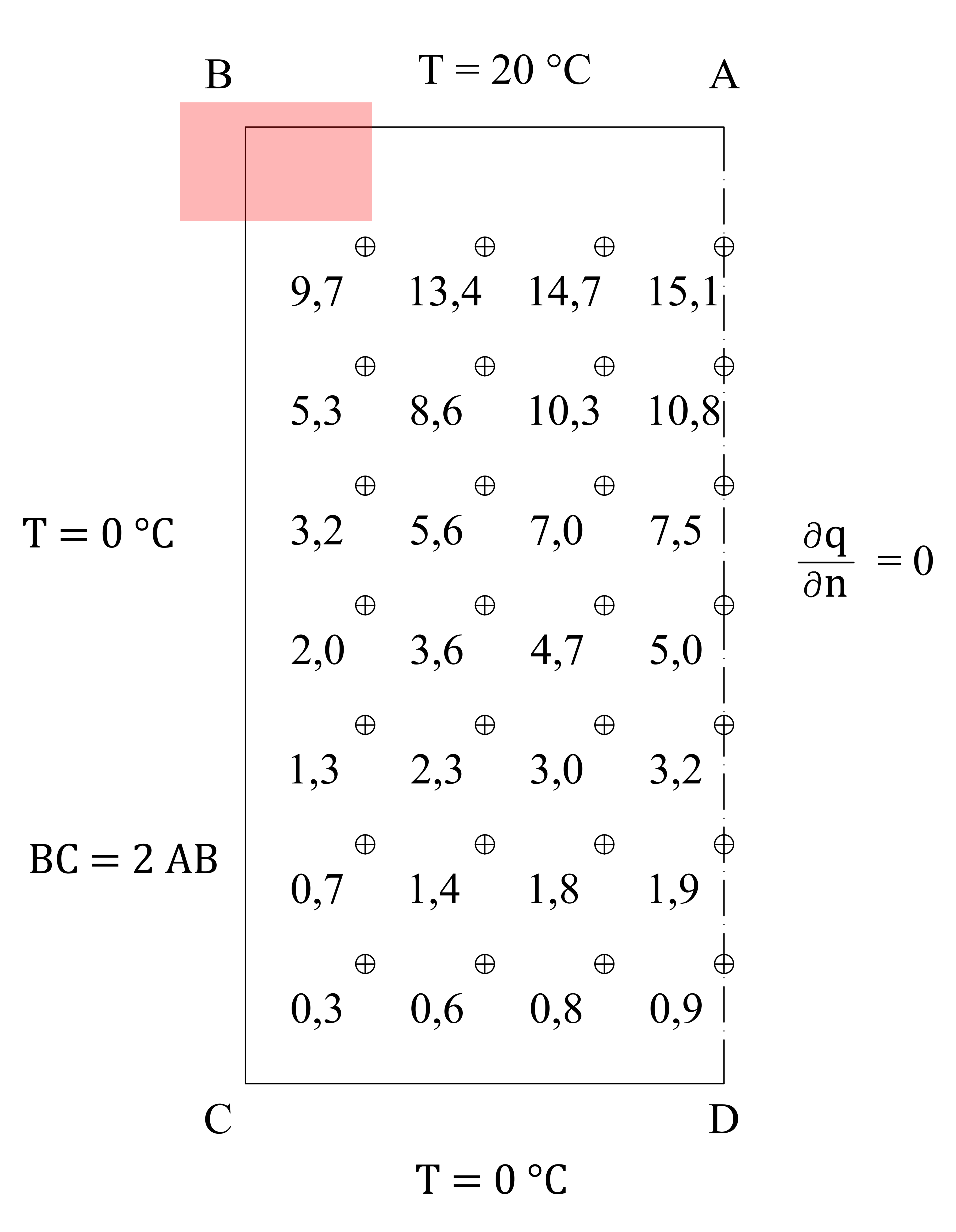}
	\caption{Case 1 from \cite{ISO}. Red area – edge with a singular point due to discretely assigned Dirichlet boundary condition.}
	\label{f1}
\end{figure}

For two and three-dimensional problems, and especially in the cases with complex geometry, the numerical solution is inevitable. To classify software as two or three-dimensional steady-state high precision method, \cite{ISO} provides a benchmark case studies. One of the requirements is that the heat flow that is entering the element should not exceed 1 \% difference between two spatial subdivisions, and the element must be subdivided until this condition is satisfied. A problem occurs for Case 1 in which the analysed problem has a singular point due to discretely assigned Dirichlet boundary conditions (Figure \ref{f1}). 

The problem shown by Case 1 is not realistic since it does not occur in buildings so one could argue by the importance of the benchmark test given by this example primarily due to the numerical instability. Finite element method (FEM) solution for the total heat flux entering the element does not converge when comparing subdivisions of 2n x 2n and n x n finite elements because temperature gradient in the marginal node in the y-direction is increasing as temperature difference remains constant and the distance between the corresponding nodes is decreasing. All the calculations carried out in this research are performed with the open-source solver for FEM Code\_Aster \cite{codeaster}.

\section{Methods}

\begin{figure}[H]
	\centering
	\includegraphics[scale=0.15]{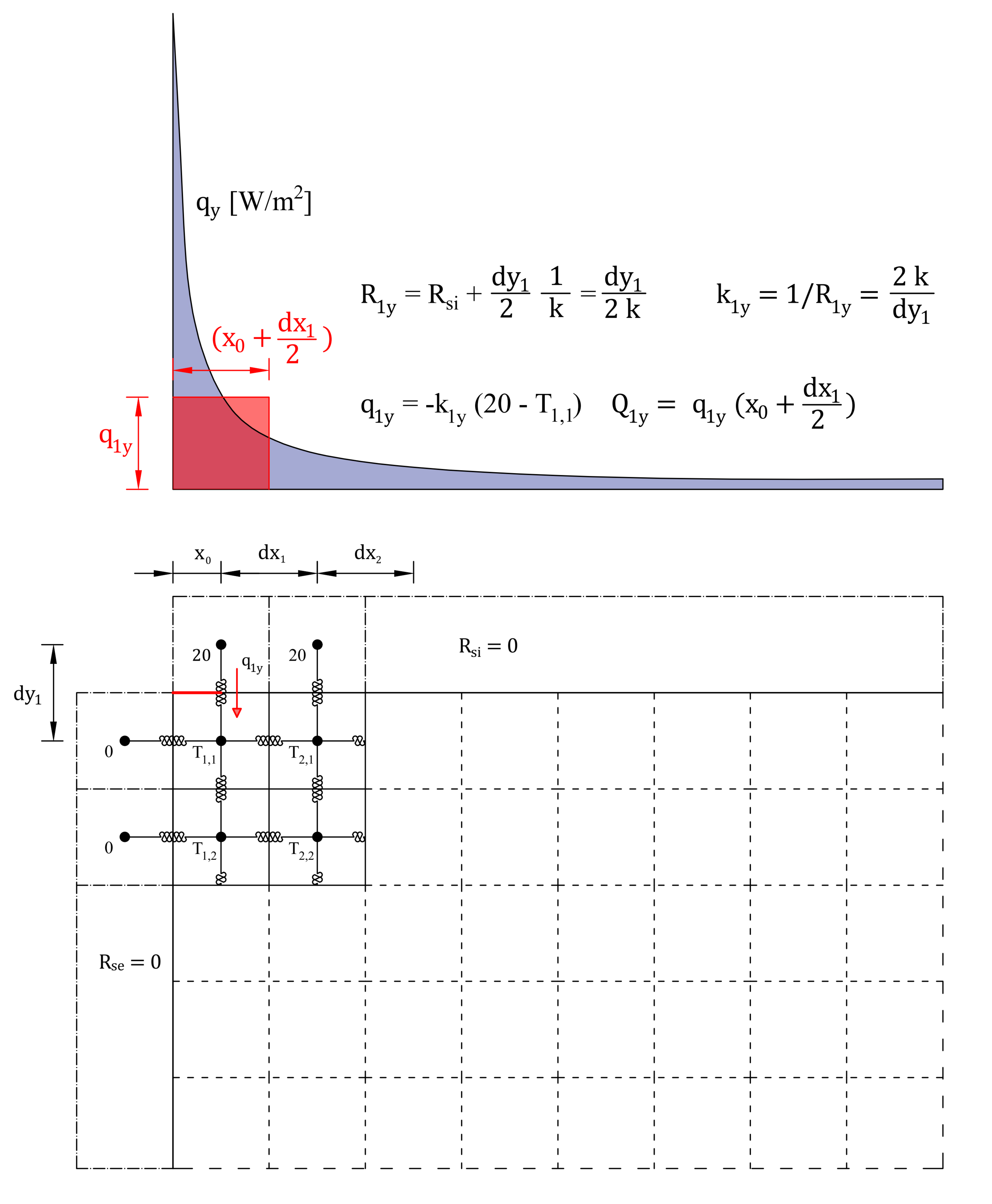}
	\caption{Control volume approach. Red line – neglected part. For marginal element, heat flux is defined for boundary temperatures and central point of the marginal element.}
	\label{f2}
\end{figure}

If the solution method of the Control Volume Method is observed (Figure \ref{f2}), then the similar principle can be applied to FEM, and the heat flow near the discretely assigned Dirichlet boundary condition can be neglected.

The main problem this research is focused on is shown in Figure \ref{f1}. The element is a rectangle with the dimensions of 0,2 x 0,4 m and an adiabatic cut-off plane along the longer side. Boundary conditions are defined according to Figure \ref{f1}. The body consists only of one material, and the thermal conductivity k is set as 1 W/(m K). 

The problem which is described with above-mentioned geometry, materials, boundary conditions and differential equation \ref{e3} is solved numerically with the FEM. For space domain discretization, linear quadrilateral elements were used, and the shape functions are defined using four nodes in the junctions of the element’s edges. As this is a thermal analysis, each node has only one degree of freedom which is the temperature of that node.

The problem is analysed for different types of subdivisions with constant element length in both directions. The subdivision of the element started with 2 × 2 cm element dimension and ended with 0.125 × 0.125 cm. In each subdivision, the element size was reduced by half. A more detailed analysis was done for the last three subdivisions.

Except for the resulting temperature field, the analysis included post-processing heat flux calculation from temperatures at Gauss points. After the calculation of heat flux is done, the heat flux results are extrapolated to each element and each node by averaging the heat flux in two adjacent nodes. The total heat flux [W/m] entering the element is observed in a more detail analysis, which is simply a heat flux in nodes on the top (with Dirichlet boundary condition 20 °C) entering the body through the effective area. The heat flux entering the element is calculated by integrating the heat flux density [W/m2] in the y-direction. The point of interest was the convergence of the total heat flux entering the element between the two subdivision of the element.

\section{Results}

\begin{figure}[H]
	\centering
	\includegraphics[scale=0.4]{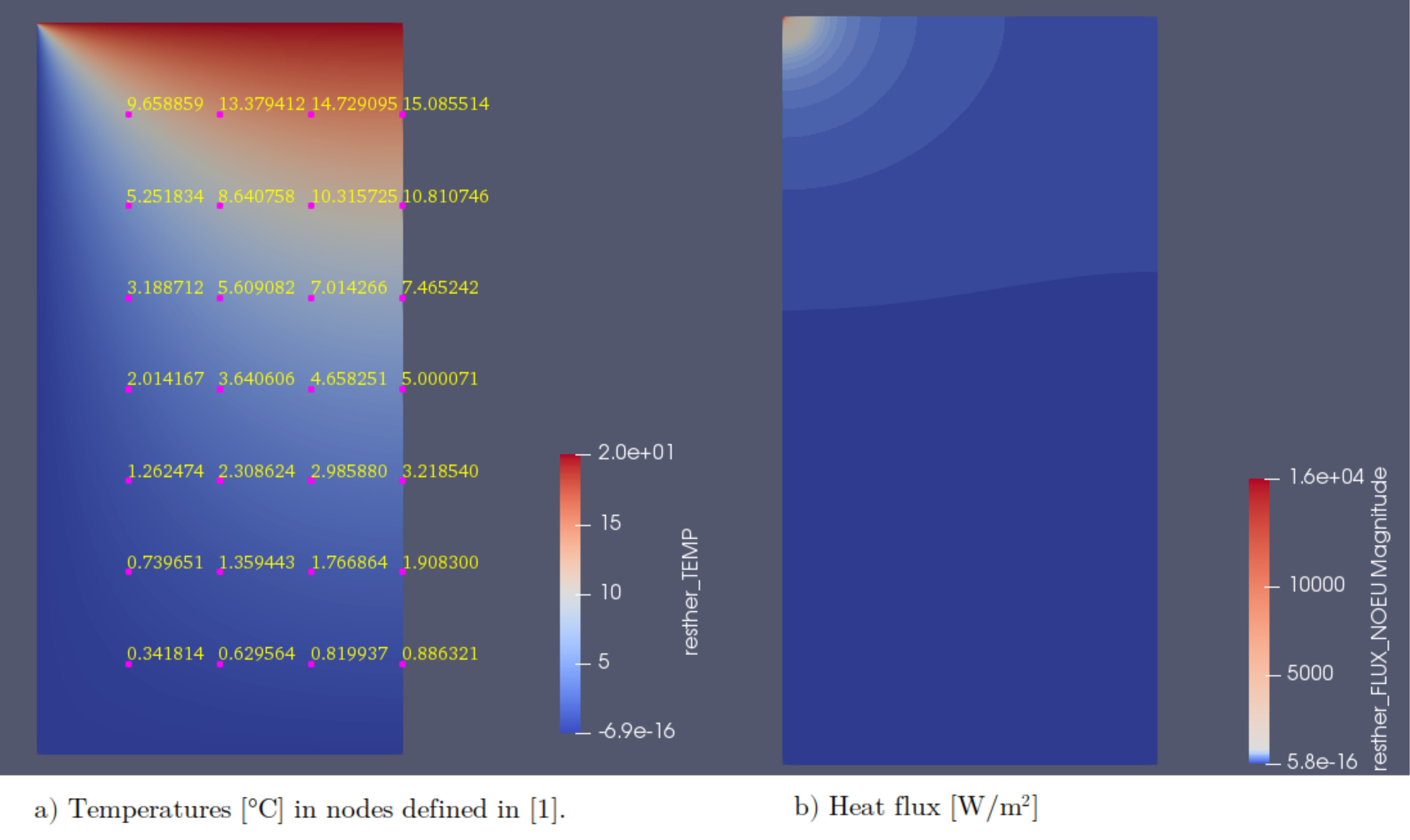}
	\caption{Results for mesh size 0.125 cm x 0.125 cm.}
	\label{f3}
\end{figure}

The results represent a linear steady-state heat transfer analysis of the temperatures in the nodes of the finite element grid from which the heat flux and the total heat flux were calculated as described in Methodology. Figure \ref{f3} shows the results for the quadrilateral 0.125 x 0.125 cm mesh. Resulting temperature field shows minimal deviation compared with the results from Figure \ref{f1}.

\begin{figure}[H]
	\centering
	\includegraphics[scale=0.4]{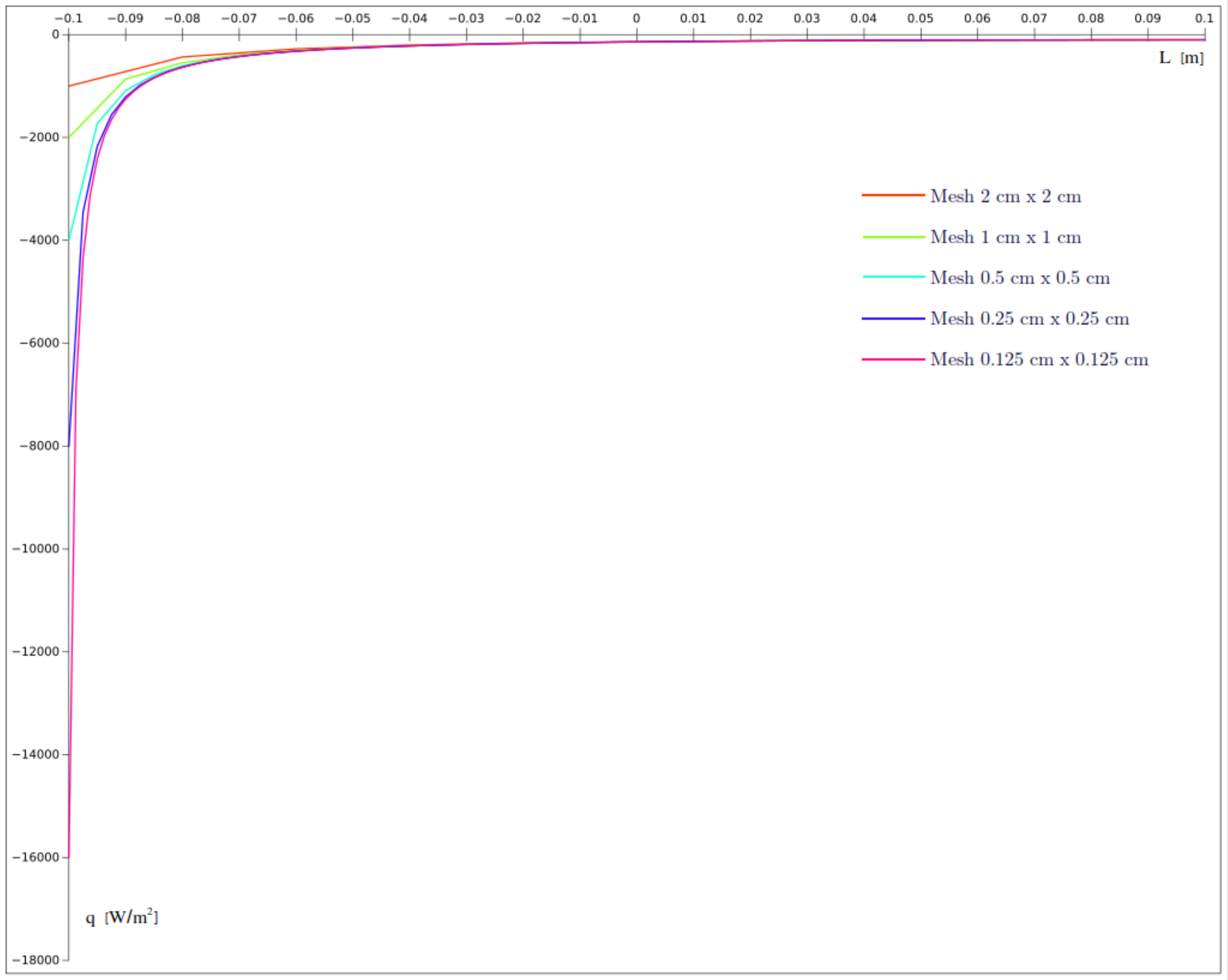}
	\caption{Heat flux entering the element in the y-direction.}
	\label{f4}
\end{figure}

In the case of the 0.125 × 0.125 cm element size, the heat flux in the marginal node is 16000 W/m2. The increase in the heat flux with the decrease of the element size is shown in Figure \ref{f4}. For marginal node the heat flux in y-direction is in the range from 1000 W/m2 for 2 cm element size (dx = dy) up to above-mentioned 16000 W/m2 for 0.125 cm element size. This is the reason why total heat flux convergence is not achieved for iterative subdivision from 2n x 2n to n x n elements.

\begin{figure}[H]
	\centering
	\includegraphics[scale=1.3]{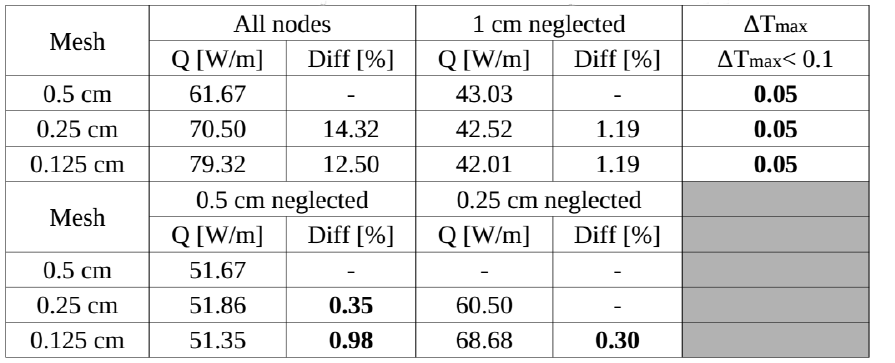}
	\caption{Results by mesh size and the comparison with \cite{ISO}.}
	\label{t1}
\end{figure}

Table \ref{t1} shows a study of three types of quadrilateral meshes. If the total heat flux in all the nodes where the heat is entering the construction detail is observed, the difference between the two subdivisions is greater than tolerance from \cite{ISO}. If this kind of subdivision is continued to the element size of 0.03125 cm, the total heat flux entering the element converges to the difference of approx. 10 \%. On the other hand, if the influence of marginal node is neglected, the convergence can be achieved because the difference for the heat flux entering the element between two subdivisions is less than 1 \%, and it decreases with decreasing of the element size. Also, for all three subdivisions, the difference between the temperatures comparing with \cite{ISO} is maximally 0.05 °C, which satisfies the tolerance of 0.1 °C.

\section{Conclusions \& Contributions}

This study was carried out to analyse the Case 1 from \cite{ISO}, which has a singular point on its boundary. To achieve the convergence of 1 \% between the total heat flux entering the element calculated for two successive subdivisions some assumptions had to be made. From the theoretical point of view, in the marginal node with the Dirichlet boundary condition of 0 °C and 20 °C, heat flux intensity approaches infinity as the element size approaches zero, which leads to unreliable results for total heat flux calculated using FEM. In reality (especially in the building sector) there is no case with this kind of boundary conditions. Even if there is a large temperature difference, there is an interval in which this change is to happen, and the heat flux has a theoretical limit which is not infinity but rather some finite number. From the results of this research, it is shown that one should neglect a certain number of singular points in order to achieve the tolerance given in \cite{ISO} since the temperature further from the marginal node is stable for any subdivision as it is shown in Figure \ref{f4}. If the heat fluxes in these marginal nodes are neglected, then the system reaches convergence (Table \ref{t1}).

This criteria for total heat flux convergence from \cite{ISO} demands that the first subdivision is defined as 2n x 2n to n x n mesh, but after this, further subdivisions are user-defined. This criteria can be satisfied by using the finite element size which is not much smaller than the previous, so the difference between two total heat fluxes is under 1 \% (e.g. for meshes 0.32 cm and 0.3125 cm, for the same case, the difference is 0.33 \%), but convergence is certainly not achieved as it can be seen in Table 1 for further subdivisions. The refinement in the area around the node with singular point does not work for this purpose as well as using the second-order serendipity elements.

\section*{Acknowledgements}

One of the authors (Sanjin Gumbarević) would like to acknowledge the Croatian Science Foundation and the European Social Fund for the support under the project ESF DOK-01-2018.


\bibliographystyle{unsrtnat}


\end{document}